\documentclass[twocolumn,pra,aps,superscriptaddress,showpacs]{revtex4}
\usepackage{mathrsfs}
\usepackage{graphicx}
\begin{document}
\title {Exact periodic and solitonic states in the spinor condensates}
\author{Zhi-Hai Zhang}
\affiliation{Department of Physics, Beijing Normal University, Beijing 100875, China}
\author{Shi-Jie Yang\footnote{Corresponding author: yangshijie@tsinghua.org.cn}}
\affiliation{Department of Physics, Beijing Normal University, Beijing 100875, China}
\affiliation{State Key Laboratory of Theoretical Physics, Institute of Theoretical Physics, Chinese Academy of Sciences, Beijing, 100190}
\begin{abstract}
We propose a method to analytically solve the one-dimensional coupled nonlinear Gross-Pitaevskii equations which govern the motion of the spinor Bose-Einstein condensates. In a uniform external potential, the Hamiltonian comprises the kinetic energy, the linear and the quadratic Zeeman energies. Several classes of exact periodic and solitonic solutions, either in real or in complex forms, are obtained for both the $F=1$ and $F=2$ condensates. These solutions are general that contain neither approximations nor constraints on the system parameters.
\end{abstract}
 \maketitle

\section{Introduction}
The experimental achievement of Stenger \emph{et al}. in trapping sodium atoms by optical means in 1998\cite{Stenger}, triggered the study of the magnetism in quantum degenerate atomic gases. Since the atom spins are not frozen, the direction of the spin can change dynamically through collisions between the atoms\cite{Ott,Gorlitz,Leanhardt}. In contrast to the scalar gases, spinor gases can host a wide variety of complex structures at zero temperature, from spin textures, magnetic crystallization, to fractional vortices \textit{et al.}\cite{Anglin,Kobayashi}. In the ground state, the symmetry is spontaneously broken in several different ways, leading to a number of possible phases\cite{Ho,Chang,Murata,Imambekov}. There exists an interplay between superfluidity and magnetism due to the spin-gauge symmetry. A ferromagnetic Bose-Einstein condensates (BECs) spontaneously creates a supercurrent as the spin is locally rotated\cite{Ueda,Kawaguchi}. The study of ultracold spinor is of primordial importance in deepening our understanding of condensed matter related issues.

The motion of the dilute spinor condensates is governed by the coupled Gross-Pitaevskii equations (GPEs). There is a large amount of works that numerically solve GPEs\cite{Li,Zhang,Nistazakis}. Analytically, some solitonic solutions are obtained by means of variable functions or similar transformation for time or spatial modulated coupling constants\cite{Beitia,Theocharis,Avelar,Wang}. Various approximations are employed to study the solitons such as bright and dark solitons in the $F=1$ spinor BECs\cite{BJ,Bradley,LLi,Yan}. Exact solutions are usually difficult to obtain due to the complexity of the coupled nonlinear GPEs. The challenges are two-fold: one is the nonlinear density-density interactions, while the other is the spin-exchange couplings between the hyperfine states. In our previous publications we have given exact solutions for the $F=1$ and $F=2$ spinor BECs for some special cases\cite{zzh,zly}. In this paper, we propose a general method which simultaneously decouples the nonlinear density-density interactions and the spin-spin interactions in the GPEs. Classes of the exact solutions, either in real or in complex forms, are systematically constructed for the Hamiltonian containing the linear and quadratic Zeeman energies. The solutions are expressed by combinations of the Jacobi elliptical functions for periodic states or the hyperbolic functions for solitonic states. The latter are identified as vector solitons or scalar solitons, respectively.

The paper is organized as follows: In Sec.II we described the method and systematically present solutions for the spin-1 condensates. In Sec.III we give a solution to the spin-2 condensates as an example. Section IV contains a brief summary.

\section{spin-1 condensates}
We are concerned with the quasi-one-dimensional (1D) spinor system in a uniform external potential ($V(\textbf{r})=0$). In this section, we deal with the $F=1$ condensates in which the meanfield order parameters are described by a macroscopic wavefunction with three hyperfine states $\Psi=(\psi_{+1},\psi_0,\psi_{-1})^T$. The Hamiltonian that contains the linear and the quadratic Zeeman effects reads\cite{Ho,Chang}
\begin{eqnarray}
H=\int d\textbf{r}&&\{\sum_{m=-1}^{1}\psi_m^*[-\frac{\hbar^2}{2M}\bigtriangledown^2+V(\textbf{r})-pm+qm^2]\psi_m\nonumber\\
&&+\frac{\bar c_0}{2}n_{tot}^2+\frac{\bar c_2}{2}|\textbf{F}|^2\},
\end{eqnarray}
where the spin-polarization vector $\textbf{F}=\psi_m^* \hat{\textbf{F}^i}_{mn}\psi_n$ with $\hat \textbf{F}^i$ ($i=x,y,z$) the spin matrices. The terms with coefficients $c_0$ and $c_2$ describe respectively the spin-independent and the spin-dependent binary elastic collisions in the combined symmetric channels of total spin $0$ and $2$. They are expressed in terms of the $s$-wave scattering lengths $a_0$ and $a_2$ as: $\bar{c}_0=4\pi\hbar^2(a_0+2a_2)/3M$ and $\bar{c}_2=4\pi\hbar^2(a_2-a_0)/3M$. $p$ and $q$ are linear and quadratic Zeeman coupling coefficients, respectively. $n_{tot}=|\psi_1|^2+\psi_0|^2+|\psi_{-1}|^2$ and $V(\textbf{r})$ is the external potential.

The dynamical motion of the $F=1$ spinor wavefunctions are governed by $i\partial_t\psi_m=\delta H/\delta\psi_m^*$, which are explicitly written as the coupled GPEs,

\begin{eqnarray}
i\partial_t\psi_m&=&[-\frac{\hbar^2}{2M}\partial_x^2-pm+qm^2+c_0n_{tot}]\psi_m\\ \nonumber
&+&c_2\sum_{n=-1}^1\textbf{F}\cdot \hat{\textbf{F}^i}\psi_n,(m=1,0,-1)\label{temporal}
\end{eqnarray}
where $c_0=\bar{c}_0/2a_{\perp}^2$ and $c_2=\bar{c}_2/2a_{\perp}^2$ are the reduced coupling constants with $a_{\perp}$ the transverse width of the quasi-1D system.

Below we choose $\hbar=M=1$ as the units for convenience. By substituting the wavefunction $\Psi(x,t)$ with
\begin{equation}
\left(
          \begin{array}{c}
          \psi_1(x,t) \\
          \psi_0(x,t) \\
          \psi_{-1}(x,t) \\
          \end{array}
        \right)
\rightarrow\left(
          \begin{array}{c}
            \psi_1(x)e^{-i(\mu+\mu_1)t}\\
            \psi_0(x)e^{-i\mu t} \\
            \psi_{-1}(x)e^{-i(\mu-\mu_1)t}\\
          \end{array}
        \right),
\end{equation}
we obtain the stationary GPEs as,
\begin{widetext}
\begin{eqnarray}
(\mu+\mu_1)\psi_1=[-\frac{1}{2}\partial_x^2+(c_0+c_2)(|\psi_1|^2+|\psi_0|^2)+(c_0-c_2)|\psi_{-1}|^2-p+q]\psi_1+c_2\psi_0^2\psi_{-1}^*,\label{stationary}\\\nonumber
\mu\psi_0=[-\frac{1}{2}\partial_x^2+(c_0+c_2)(|\psi_1|^2+|\psi_{-1}|^2)+c_0|\psi_0|^2]\psi_0+2c_2\psi_0^*\psi_1\psi_{-1},\\\nonumber
(\mu-\mu_1)\psi_{-1}=[-\frac{1}{2}\partial_x^2+(c_0+c_2)(|\psi_{-1}|^2+|\psi_0|^2)+(c_0-c_2)|\psi_1|^2+p+q]\psi_{-1}+c_2\psi_0^2\psi_1^*.
\end{eqnarray}
\end{widetext}
Since the chemical potential of each hyperfine state is different, Eqs.(\ref{stationary}) are not really stationary equations given $\mu_1\neq 0$. It comprises a "Lamor precession" between the hyperfine states. However, the density profile of each state are time-invariant so we simply call them the stationary states. The periodic boundary conditions,
\begin{equation}
\psi_m(1)=\psi_m(0),\hspace{5mm}
\psi_m'(1)=\psi_m'(0).\label{boundary}
\end{equation}
is adopted in our calculations. We consider several types of real and complex solutions for $F=1$ condensates.

In order to seek the analytical solutions we decouple spin-spin interactions in the Eqs.(\ref{stationary}) by requiring $\psi_{-m}^*=\pm\psi_m$ and $\psi_0^*=\psi_0$. It is directly to check that $\psi_{-m}^*=\psi_m$ corresponds to (partially) spin-polarized states ($|\textbf{F}|\neq 0$) whereas $\psi_{-m}^*=-\psi_m$ corresponds to spin-unpolarized or polar states ($|\textbf{F}|= 0$). On the other hand, the nonlinear density-density couplings between the hyperfine states are decoupled by making use of the properties of the Jacobi elliptical functions or the hyperbolic functions. There are real and complex forms of solutions which are explicitly described as follows.

\subsection{Real solutions}
We first consider the following $\textrm{sn}$-$\textrm{cn}$-$\textrm{sn}$ form of solution to the nonlinear Eq.(\ref{stationary}),
\begin{equation}
\left(\begin{array}{l}\psi_1(x)\\
\psi_0(x)\\ \psi_{-1}(x)
\end{array}\right)
=\left(\begin{array}{l}A \textrm{sn}(kx,m)\\
D \textrm{cn}(kx,m)\\ - A \textrm{sn}(kx,m)
\end{array}\right),\label{real solution}
\end{equation}
where $\textrm{sn}$ and $\textrm{cn}$ are the Jacobi elliptical functions of modulus $m$. The period is $k=4jK(m)$ with $j$ the number of periods which will always be set to $j=2$ in the figures). $A$ and $D$ are the real constants. From the relation $\textrm{sn}^2+\textrm{cn}^2=1$, one has
\begin{equation}
|\psi_{-1}|^2=|\psi_{1}|^2,\hspace{3mm}
|\psi_0|^2=D^2-\frac{D^2}{A^2}|\psi_1|^2.\label{real}
\end{equation}
By substituting (\ref{real}) into the equations (\ref{stationary}), we obtain three decoupled equations
\begin{equation}
\tilde{\mu}_m\psi_m=-\frac{1}{2}\psi_m^{\prime\prime}+\tilde{\gamma}_m|\psi_m|^2\psi_m,\hspace{3mm}(m=0,
\pm 1) \label{decouple}
\end{equation}
where the effective chemical potentials $\tilde\mu_m$ and interacting constants $\tilde\gamma_m$ are defined as
\begin{equation}
\left.\begin{array}{l}\tilde{\mu}_{\pm 1}=\mu-q-c_0D^2,\\
\tilde{\mu}_0=\mu-2c_0A^2,\\
\tilde{\gamma}_{\pm 1}=2c_0-c_0\frac{D^2}{A^2},\\
\tilde{\gamma}_0=c_0-2c_0\frac{A^2}{D^2}.\label{rel1}
\end{array}\right.
\end{equation}
\begin{figure}
\begin{center}

\includegraphics*[width=9cm]{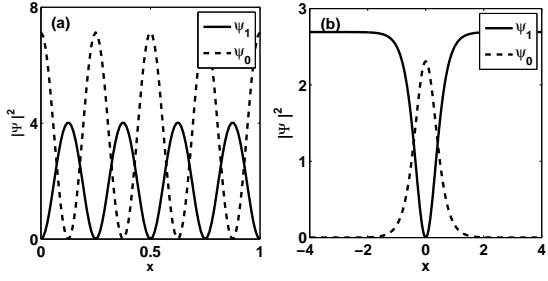}
\caption{(a) The density profiles of the solution (\ref{real solution}). $c_0=30$, $q=-13.7707$, $m=0.4$ and $\mu=300$. (b) The density profiles for the single soliton solution (\ref{real soliton}). $c_0=1.3$, $k=2$, $q=-2$ and $\mu=5$.}
\end{center}
\end{figure}

In order to obtain self-consistent solutions, one should take $p=-\mu_1$. Therefore, the linear Zeeman energy plays the role of balancing the chemical potentials between the hyperfine states $\psi_1$ and $\psi_{-1}$. The effective chemical potentials $\tilde\mu_m$ and amplitudes $A$, $D$ can be obtained as,
\begin{equation}
\left.\begin{array}{l}\tilde{\mu}_{\pm
1}=\frac{1}{2}k^2(1+m^2),\\
\tilde{\mu}_{0}=\frac{1}{2}k^2(1-2m^2),\\
A^2=\frac{m^2k^2}{\tilde{\gamma}_{\pm 1}},\\
D^2=-\frac{m^2k^2}{\tilde{\gamma}_0}.\label{real result}
\end{array}\right.
\end{equation}
From relations (\ref{rel1}) and (\ref{real result}) we conclude that the effective intra-species interactions for hyperfine states $\psi_{\pm}$ should be repulsive ($\tilde\gamma_{\pm 1}>0$) while for $\psi_0$ be attractive ($\tilde\gamma_0<0$). These impose constraint relations on the values of parameters $c_0$, $q$ and $\mu$.

Figure 1(a) illustrates the density profiles of each hyperfine states for solution (\ref{real solution}). The parameters are chosen as $c_0=30$, $q=-13.7707$, $A=2.0061$, $D=2.6704$, $m=0.4$ and $\mu=300$. This state has vanishing spin-polarization $|\textbf{F}|=0$. The modulus of the Jacobi elliptical functions is a free parameter. As $m\rightarrow 1$, we naturally obtain the periodic soliton train solution. The single soliton  solution can be obtained by directly substituting the Jacobi elliptical functions with the hyperbolic functions in (\ref{real solution}). Namely,
\begin{equation}
\left(\begin{array}{l}\psi_1(x)\\
\psi_0(x)\\ \psi_{-1}(x)
\end{array}\right)
=\left(\begin{array}{l}A \tanh(kx)\\
D \textrm{sech}(kx)\\ - A \tanh(kx)
\end{array}\right),\label{real soliton}
\end{equation}
This solution has been addressed in literatures\cite{Li,Nistazakis}. The density profile is displayed in Fig.1(b), with the parameters $c_0=1.3$ ,$k=2$, $q=-2$, $A=1.6408$, $D=1.5191$ and $\mu=5$. It is a dark-bright-dark composite soliton.

Other forms of real solutions can also be constructed by the same way. For example, we seek the  $\textrm{cn}$-$\textrm{sn}$-$\textrm{cn}$ form of solution to Eq.(\ref{stationary}),
\begin{equation}
\left(\begin{array}{l}\psi_1(x)\\
\psi_0(x)\\ \psi_{-1}(x)
\end{array}\right)
=\left(\begin{array}{l}
A \textrm{cn}(kx,m)\\
D \textrm{sn}(kx,m)\\
- A \textrm{cn}(kx,m)
\end{array}\right).\label{real solution1}
\end{equation}
One has
\begin{equation}
|\psi_{-1}|^2=|\psi_{1}|^2,\hspace{3mm}
|\psi_0|^2=D^2-\frac{D^2}{A^2}|\psi_1|^2.\label{real1}
\end{equation}
Eq.(\ref{stationary}) are again decoupled into (\ref{decouple}). It follows that the effective chemical potentials $\tilde\mu_m$ and interacting constants $\tilde\gamma_m$,
\begin{equation}
\left.\begin{array}{l}
\tilde{\mu}_{\pm 1}=\mu-q-c_0D^2,\\
\tilde{\mu}_0=\mu-2c_0A^2,\\
\tilde{\gamma}_{\pm 1}=2c_0-c_0\frac{D^2}{A^2},\\
\tilde{\gamma}_0=c_0-2c_0\frac{A^2}{D^2}.\label{rel2}
\end{array}\right.
\end{equation}
The effective chemical potentials $\tilde\mu_m$ and amplitudes $A$, $D$ are self-consistently calculated as,
\begin{equation}
\left.\begin{array}{l}
\tilde{\mu}_{\pm1}=\frac{1}{2}k^2(1-2m^2),\\
\tilde{\mu}_{0}=\frac{1}{2}k^2(1+m^2),\\
A^2=-\frac{m^2k^2}{\tilde{\gamma}_{\pm 1}},\\
D^2=\frac{m^2k^2}{\tilde{\gamma}_0}.\label{real result1}
\end{array}\right.
\end{equation}

\begin{figure}
\begin{center}
\includegraphics*[width=9cm]{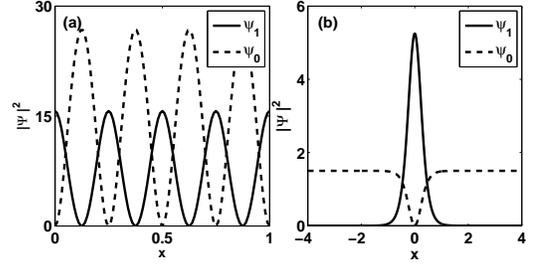}
\caption{(a) The density profiles of the solution (\ref{real solution1}) with $c_0=-10$, $q=22.734$, $m=0.5$ and $\mu=-200$. (b) The density profiles for the the single soliton solution corresponding to periodic solution (\ref{real solution1}) with $c_0=-1$, $k=3$, $q=4.5$ and $\mu=-1.5$.}
\end{center}
\end{figure}

Relations (\ref{rel2}) and (\ref{real result1}) reveal that the effective intra-species interactions for the hyperfine states $\psi_{\pm}$ should be attractive ($\tilde\gamma_{\pm 1}<0$) while for $\psi_0$ be repulsive ($\tilde\gamma_0>0$).

Figure 2(a) plots the density profiles of each hyperfine states for solution (\ref{real solution1}). The parameters are $c_0=-10$, $q=22.734$, $A=3.9602$, $D=5.1788$, $m=0.5$ and $\mu=-200$. In this state, the spin is unpolarized $|\textbf{F}|=0$. Similarly, the single solitonic solution is constructed by substitution $\textrm{sn}\longrightarrow \tanh$ and $\textrm{cn}\longrightarrow \textrm{sech}$. Figure 2(b) is the density profiles of a typical bright-dark-bright soliton with $c_0=-1$, $k=3$, $q=4.5$, $A=2.2913$, $D=1.2247$ and $\mu=-1.5$.

\subsection{Complex solutions}
We seek complex forms of periodic solution to the nonlinear Eq.(\ref{stationary}) as,
\begin{equation}
\left(\begin{array}{l}
\psi_{1}\\
\psi_0\\
\psi_{-1}
\end{array}\right)
=\left(\begin{array}{l}
f(x)e^{i\theta(x)}\\
D \textrm{sn}(kx,m)\\
f(x)e^{-i\theta(x)}
\end{array}\right),\label{complex solution}
\end{equation}
where $f(x)=\sqrt{A+B \textrm{cn}^2(kx,m)}$. $A$, $B$ and $D$ are real constants. One has
\begin{equation}
|\psi_{-1}|^2=|\psi_1|^2, \hspace{3mm}
|\psi_0|^2=D^2(1+\frac{A}{B})-\frac{D^2}{B}|\psi_1|^2.\label{complex}
\end{equation}
By substituting (\ref{complex}) into the coupled GPEs (\ref{stationary}), we again obtain the decoupled equations (\ref{decouple}) with the effective chemical potentials and intr-species interaction constants,
\begin{equation}
\left.\begin{array}{l}
\tilde{\mu}_{\pm 1}=\mu-q-(c_0+2c_2)\frac{D^2}{B}(A+B),\\
\tilde{\mu}_0=\mu-2(c_0+2c_2)(A+B),\\
\tilde{\gamma}_{\pm 1}=2c_0-(c_0+2c_2)\frac{D^2}{B},\\
\tilde{\gamma}_0=c_0-2(c_0+2c_2)\frac{B}{D^2}.\\
\end{array}\right.
\end{equation}

In order to obtain the self-consistent solution, we set $\mu_1=-p$ which yields to
\begin{equation}
\left.\begin{array}{l}
B=-\frac{m^2 k^2}{\tilde\gamma_1},\\
A=\frac{2\tilde\mu_{\pm 1}-(1-2m^2)k^2}{3\tilde\gamma_{\pm 1}},\\
D^2=\frac{m^2k^2}{\tilde\gamma_0}.
\end{array}\right.
\end{equation}

We note that the effective interactions $\tilde\gamma_0>0$. The phase is
\begin{equation}
\theta(x)=\int^x_0\frac{\alpha}{f^2(\xi)}d\xi,
\end{equation}
where $\alpha=\pm (2\tilde\mu_{\pm 1} A^2-2\tilde\gamma_{\pm 1} A^3+k^2AB (1-m^2))^{\frac{1}{2}}$
is an integral constant. The periodic boundary conditions (\ref{boundary}) require that the amplitude and phase satisfy, respectively, 
\begin{equation}
f(1)=f(0), \hspace{3mm} \theta(1)-\theta(0)=2j\pi\times n,\label{boundary1}
\end{equation}
where $n$ is an integer. The periodic condition for the phase can be fulfilled by properly adjusting the modulus $m$ of the Jacobi elliptical functions. Figure 3(a) and (b) display the phase and density profiles of the complex solution (\ref{complex solution}), respectively. The parameters are taken as $n=2$, $c_0=34$, $c_2=43$, $q=17$, $A=1.524$, $B=-0.5795$, $D=1.0766$, $m=0.82$ and $\mu=448.6349$. As $m\rightarrow 1$, it results into the soliton train state.

\begin{figure}
\begin{center}
\includegraphics*[width=9cm]{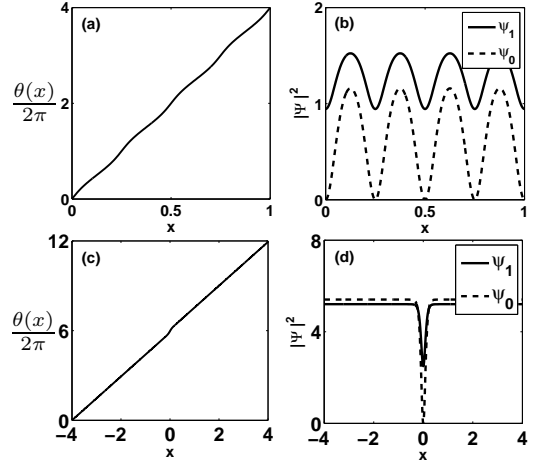}
\caption{The phase profile (a) and the density profile (b) for the complex solutions (\ref{complex solution}). The parameters are $n=2$, $m=0.82$, $q=17$, and $\mu=448.6349$. The phase profile (c) and the density profile (d) for the single soliton solution (\ref{complex soliton}) with $n=6$, $k=8.7$, $q=5$, and $\mu=130.69$.}
\end{center}
\end{figure}

The single soliton solution is obtained by substituting the Jacobi elliptical functions with the hyperbolic functions as
\begin{equation}
\left(\begin{array}{l}
\psi_{1}\\
\psi_0\\
\psi_{-1}
\end{array}\right)
=\left(\begin{array}{l}
f(x)e^{i\theta(x)}\\
D \tanh(kx)\\
f(x)e^{-i\theta(x)}
\end{array}\right),\label{complex soliton}
\end{equation}
where $f(x)=\sqrt{A+B \textrm{sech}^2(kx)}$. Figure 3(c) and (d) plots a typical grey-dark-grey soliton for the solution (\ref{complex soliton}) with $n=6$, $c_0=3$, $c_2=4$, $q=5$, $k=8.7$, $A=5.2032$, $B=-2.7032$, $D=2.3252$ and $\mu=130.69$. Here the phases $\theta(x)$ should satisfy the periodic condition (\ref{boundary}) by properly adjusting the width of the soliton $k$. Since ($|\textbf{F}|\neq 0$), this type of composite soliton may be properly called the polarized or vector soliton.

The spin-unpolarized ($\textbf{F}=0$) complex solution can be constructed as,
\begin{equation}
\left(\begin{array}{l}
\psi_{1}\\
\psi_0\\
\psi_{-1}
\end{array}\right)
=\left(\begin{array}{l}
f(x)e^{i\theta(x)}\\
D \textrm{cn}(kx,m)\\
-f(x)e^{-i\theta(x)}
\end{array}\right),\label{complex solution1}
\end{equation}
where $f(x)=\sqrt{A+B \textrm{sn}^2(kx,m)}$. $A$, $B$ and $D$ are real
constants. One has
\begin{equation}
|\psi_{-1}|^2=|\psi_1|^2, \hspace{3mm}
|\psi_0|^2=D^2(1+\frac{A}{B})-\frac{D^2}{B}|\psi_1|^2.\label{complex1}
\end{equation}
The effective chemical potentials and interaction constants are,
\begin{equation}
\left.\begin{array}{l}
\tilde{\mu}_{\pm 1}=\mu-q-c_0\frac{D^2}{B}(A+B),\\
\tilde{\mu}_0=\mu-2c_0(A+B),\\
\tilde{\gamma}_{\pm 1}=2c_0-c_0\frac{D^2}{B},\\
\tilde{\gamma}_0=c_0-2c_0\frac{B}{D^2}
\end{array}\right.
\end{equation}
One obtains
\begin{equation}
\left.\begin{array}{l}
B=\frac{m^2 k^2}{\tilde\gamma_1},\\
A=\frac{2\tilde\mu_{\pm 1}-(1+m^2)k^2}{3\tilde\gamma_{\pm 1}},\\
D^2=-\frac{m^2k^2}{\tilde\gamma_0},
\end{array}\right.
\end{equation}
which require that the effective intra-species interactions be attractive ($\tilde\gamma_0<0$).

\begin{figure}
\begin{center}
\includegraphics*[width=9cm]{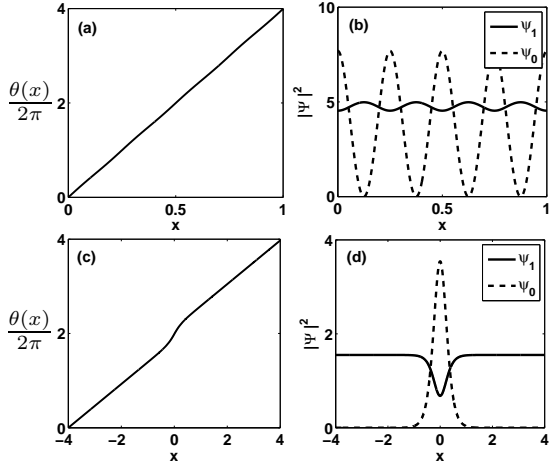}
\caption{The phase (a) and the density (b) distributions of complex solutions (\ref{complex solution1})
for the $F=1$ condensates with $n=2$, $m=0.44$, $q=-190$, $\mu=4$. (c) and (d) are the phase and density distributions for the single soliton solutions of (\ref{complex solution1}) with $n=2$, $k=3$, $q=-8$, and $\mu=-20$.}
\end{center}
\end{figure}

Figure 4(a) and (b) display the phase and density profiles of the solution (\ref{complex solution1}). The parameters are $n=2$, $c_0=-5$, $q=-190$, $A=4.5347$, $B=0.4455$, $D=2.7731$, $m=0.44$ and $\mu=4$. Figure 4(c) and (d) are the corresponding single soliton solution with parameters $n=2$, $c_0=-5$, $q=-8$, $k=3$, $A=0.6781$, $B=0.8719$, $D=1.8825$ and $\mu=-20$.  It is a typical grey-bright-grey composite soliton. In comparison to the polarized soliton (\ref{complex soliton}), we may call this spin-unpolarized soliton the polar or scalar solitons.

\section{spin-2 condensates}
Our method is also applicable to the $F=2$ condensates. For illustration, we give an example in this section. By substituting the wavefunction with\cite{Ueda,zly}
\begin{equation}
\left(
          \begin{array}{c}
          \psi_2(x,t) \\
          \psi_1(x,t) \\
          \psi_0(x,t) \\
          \psi_{-1}(x,t) \\
          \psi_{-2}(x,t) \\
          \end{array}
        \right)
\rightarrow\left(
          \begin{array}{c}
            \psi_2(x)e^{-i(\mu+\mu_2)t} \\
            \psi_1(x)e^{-i(\mu+\mu_1)t}\\
            \psi_0(x)e^{-i\mu t} \\
            \psi_{-1}(x)e^{-i(\mu-\mu_1)t}\\
            \psi_{-2}(x)e^{-i(\mu-\mu_2)t} \\
          \end{array}
        \right),
\end{equation}
we obtain the generalized stationary GPEs as,
\begin{eqnarray}
(\mu \pm \mu_2) \psi_{\pm2}&=&[-\frac{1}{2}\partial_x^2+c_0 n_{tot} \pm 2c_1 F_z \mp 2p+4q]\psi_{\pm2}\nonumber\\
&&+c_1F_{\mp} \psi_{\pm1}+\frac{c_2}{\sqrt{5}}A\psi_{\mp2}^*,\nonumber\\
(\mu \pm \mu_1) \psi_{\pm1}&=&[-\frac{1}{2}\partial_x^2+c_0 n_{tot} \pm c_1 F_z\mp p+q]\psi_{\pm1}\nonumber\\
&&+c_1(\frac{\sqrt{6}}{2}F_{\mp}\psi_0+F_{\pm}\psi_{\pm2})-\frac{c_2}{\sqrt{5}}A\psi_{\mp1}^*,\nonumber\\
\mu \psi_0&=&[-\frac{1}{2}\partial_x^2+c_0 n_{tot}]\psi_0+\frac{\sqrt{6}}{2}c_1(F_+\psi_1+F_-\psi_{-1})\nonumber\\
&&+\frac{c_2}{\sqrt{5}}A\psi_0^*.\nonumber\\
\label{stationary2}
\end{eqnarray}

We seek the following form of solution which satisfies $\psi_{-m}^*=(-1)^m\psi_m$ and $\psi_0^*=\psi_0$,
\begin{equation}
\left(\begin{array}{l}
\psi_{2}\\
\psi_{1}\\
\psi_0\\
\psi_{-1}\\
\psi_{-2}
\end{array}\right)
=\left(\begin{array}{l}
f(x)e^{i\theta(x)}\\
C \textrm{sn}(kx,m)\\
D \textrm{cn}(kx,m)\\
-C \textrm{sn}(kx,m)\\
f(x)e^{-i\theta(x)}
\end{array}\right),\label{complex solution2}
\end{equation}
where $f(x)=\sqrt{A+B \textrm{sn}^2(kx,m)}$. $A$, $B$, $C$ and $D$ are real constants. This form of solution has vanishing spin-polarization $|{\bf F}|=0$. The equations (\ref{complex solution2}) can be decoupled and self-consistently solved in the same way as for the $F=1$ condensates. The linear Zeeman energy satisfies $\mu_1=-p$ and $\mu_2=2p$ so as to balancing the chemical potential differences between the hyperfine states. We skip the calculation details and just illustrate the results.

\begin{figure}
\begin{center}
\includegraphics*[width=9cm]{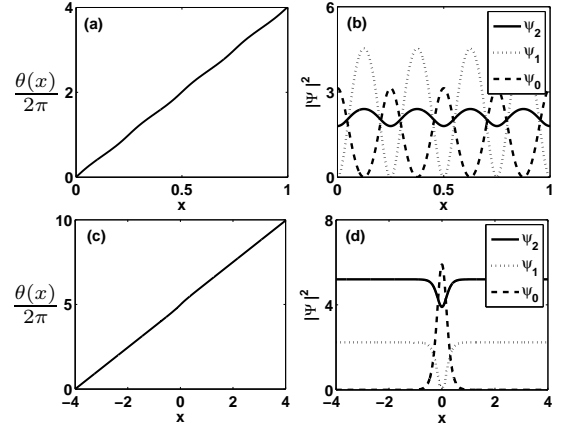}
\caption{The profiles of phase (a) and the density (b) for $F=2$ BEC solution (\ref{complex solution2}). The parameters are $n=2$, $m=0.7$, $q=-53.4153$, $\mu=210$. The profiles of phase (c) and density (d) for the  single soliton solution corresponding to (\ref{complex solution2}) with $n=5$, $k=4.05$, $q=-8.2012$, and $\mu=200$.}
\end{center}
\end{figure}

Figure 5(a) and (b) display the phase and density profiles of the solution (\ref{complex solution2}) for $F=2$ condensates. The parameters are $n=2$, $c_0=10$, $c_2=25$, $q=-53.4153$, $m=0.7$, $A=1.8$, $B=0.6$, $C=2.1278$, $D=1.7699$ and $\mu=210$. Figure 5(c) and (d) are results for the single soliton solution which corresponding to the periodic solution (\ref{complex solution2}) with $n=5$, $c_0=10$, $c_2=20$, $q=-8.2012$, $k=4.05$, $A=3.9$, $B=1.3$, $C=1.4952$, $D=2.429$ and $\mu=200$. It exhibits a typical grey-dark-bright-dark-grey composite soliton structure.

\section{summary}
We have systematically solved the one-dimensional coupled nonlinear GPEs which govern the motion of the spinor BECs exposed in a uniform magnetic field. Both periodic and solitonic stationary solutions for the $F=1$ and $F=2$ condensates are constructed. Other forms of solutions with different combinations of the Jacobi elliptical functions or hyperbolic functions can also be obtained in the same way. Our method is general and exact, without any approximations or special constraints on the system parameters. It may be extended to other nonlinear systems such as the coupled nonlinear Klein-Gordon equations or the dynamical coupled nonlinear Schrodinger equations.

This work is supported by the funds from the Ministry of Science and Technology of China under Grant No. 2012CB821403.

\end{document}